\documentclass[sigconf]{acmart}

\usepackage{subcaption}
\usepackage{dsfont}
\usepackage{booktabs}
\usepackage{xcolor}
\usepackage{pifont}   % in preamble
\usepackage[acronym]{glossaries}
\usepackage{tikz}
\usepackage{caption}
\usepackage{listings}
\usepackage{tabularx}
\usepackage{enumitem}
\usepackage{float}
\usepackage{pgfplots}
\pgfplotsset{compat=1.18}
\usepackage{xcolor}
\definecolor{pastelblue}{RGB}{114,147,203}
\definecolor{pastelpink}{RGB}{225,151,151}

\newenvironment{compact_enum}{
  \begin{itemize}[leftmargin=*, itemsep=0pt, parsep=2pt, topsep=2pt]
}{
  \end{itemize}
}

\usepackage{calc}
\newlength{\redactwidth}

\usepackage[most]{tcolorbox}
\tcbset{
  myblock/.style={
    enhanced,
    colback=gray!10,
    colframe=black,
    boxsep=5pt,
    arc=2pt,
    left=6pt,right=6pt,top=6pt,bottom=6pt
  }
}

\newcommand{\partialmark}{%
  \tikz[baseline=-0.6ex, scale=0.9]{
    \node[text=black, inner sep=0pt] (t) {\ding{51}};
    \draw[black, very thick] (-0.08,0.08) -- (0.08,-0.08); 
  }%
}
\makeglossaries
\newacronym{rl}{RL}{Reinforcement Learning}
\newacronym{llm}{LLM}{Large Language Model}
\newacronym{ppo}{PPO}{proximal policy optimization}
\newacronym{grpo}{GRPO}{Group relative policy optimization}
\newacronym{dapo}{DAPO}{decoupled clip and dynamic sampling policy optimization}
\newacronym{sft}{SFT}{Supervised Fine-Tuning}
\newacronym{ndcg}{NDCG}{normalized discounted cumulative gain}
\newacronym{verl}{VeRL}{Volcano Engine Reinforcement Learning for LLMs}
\newacronym{lora}{LoRA}{Low-Rank Adaptation}
\newacronym{recsys}{RecSys}{recommender systems}

\usepackage[most]{tcolorbox}
\tcbset{
  colback=gray!5!white,
  boxrule=0.5pt,
  left=6pt, right=6pt, top=6pt, bottom=6pt
}
\usepackage[table]{xcolor}


\title{High Fidelity Textual User Representation over Heterogeneous Sources via Reinforcement Learning}

\begin{document}

\settopmatter{authorsperrow=4} % try 3–5 depending on names
\author{Rajat Arora}
\authornote{These authors contributed equally to this research.}
\email{rajarora@linkedin.com}
\affiliation{%
  \institution{LinkedIn Corporation}
  \country{USA}
}

\author{Ye Tao}
\authornotemark[1] % co-first author note
\authornote{Work done during an internship at LinkedIn Corporation, Sunnyvale, USA}
\email{yt371@rutgers.edu}
\affiliation{%
  \institution{Rutgers, The State University of New Jersey}
  \country{USA}
}

\author{Jianqiang Shen}
\email{jshen@linkedin.com}
\affiliation{%
  \institution{LinkedIn Corporation}
  \country{USA}
}

\author{Ping Liu}
\email{piliu@linkedin.com}
\affiliation{%
  \institution{LinkedIn Corporation}
  \country{USA}
}

\author{Muchen Wu}
\email{muwu@linkedin.com}
\affiliation{%
  \institution{LinkedIn Corporation}
  \country{USA}
}

\author{Qianqi Shen}
\email{qishen@linkedin.com}
\affiliation{
 \institution{LinkedIn Corporation}
 \city{}
 \state{}
 \country{USA}
}

\author{Benjamin Le}
\email{ble@linkedin.com}
\affiliation{%
  \institution{LinkedIn Corporation}
  \country{USA}
}

\author{Fedor Borisyuk}
\email{fborisyuk@linkedin.com}
\affiliation{%
  \institution{LinkedIn Corporation}
  \country{USA}
}

\author{Jingwei Wu}
\email{jingwu@linkedin.com}
\affiliation{
 \institution{LinkedIn Corporation}
 \city{}
 \state{}
 \country{USA}
}

\author{Wenjing Zhang}
\email{wzhang@linkedin.com}
\affiliation{
 \institution{LinkedIn Corporation}
 \city{}
 \state{}
 \country{USA}
}

\renewcommand{\shortauthors}{Arora*, Tao* et al.}
\newcommand{\notepl}[1]{{\color{magenta}PL: #1}}
\newcommand{\noteyt}[1]{{\color{blue}YT: #1}}

% Short running head

% Move emails out of the main block (optional but makes it tighter)

\begin{abstract}
% LinkedIn is the world's leading jobs platform, connecting millions of Daily Active Users with jobs that best fit their profiles and interests. To build effective and personalized search and recommendation products, we need to contextualize information across multiple sources of member data such as User Profile, Resume, Job Engagement Data, and Search Queries. With the rising adoption of LLMs throughout our ranking and retrieval stack, it is also important that these sources can be easily interfaced with transformer-based LLMs. In this work, we propose to learn a single textual representation for a member across multiple input domains through Reinforcement Learning, which is not only easy to use but also interpretable. In addition, since online systems still rely on LLMs with lower parameter counts (<3B), it is important that the representation is concise and non-redundant. We show that engagement logs alone can act as useful reward signals, and that when combined with simple rule-based rewards to control output format and length, they lead to effective textual representations that are useful for multiple downstream products.
Effective personalization on large-scale job platforms requires modeling members based on heterogeneous textual sources, including profiles, professional data, and search activity logs. As recommender systems increasingly adopt Large Language Models (LLMs), creating unified, interpretable, and concise representations from heterogeneous sources becomes critical, especially for latency-sensitive online environments. In this work, we propose a novel Reinforcement Learning (RL) framework to synthesize a unified textual representation for each member. Our approach leverages implicit user engagement signals (e.g., clicks, applies) as the primary reward to distill salient information. Additionally, the framework is complemented by rule-based rewards that enforce formatting and length constraints. Extensive offline experiments across multiple LinkedIn products, one of the world's largest job platforms, demonstrate significant improvements in key downstream business metrics. This work provides a practical, labeling-free, and scalable solution for constructing interpretable user representations that are directly compatible with LLM-based systems.
\end{abstract}


\maketitle

\section{Introduction}

LinkedIn, the world's leading professional network, serves 70 million active job seekers weekly, connecting them with relevant opportunities through product features such as job search, recommendations, and alerts. At this scale, effective personalization hinges on creating a unified and powerful representation of each member for downstream models responsible for retrieval and ranking. The key challenge lies in synthesizing information from diverse and often overlapping sources, such as member profiles, professional records, as well as different types of relevant job search activity on the platform.

% Conventionally, such representations are obtained via dense embeddings and/or sparse features. However, dense representations are hard to interpret and require extensive downstream finetuning whenever the embedding generation backbone changes. Furthermore, the rapid advancement in \gls{llm} capabilities and the corresponding decrease in compute costs are leading to their increasing adoption as core retrieval and ranking engines. Conventional Embeddings and sparse features are not only incompatible for direct use with \gls{llm}s but also hinder iteration velocity, since any model needs to be continually finetuned to work with existing embeddings and features. Although Semantic IDs~\cite{singh2024better} offer the benefit of partial interpretability and compatibility with \gls{llm}s, they still require extensive finetuning~\cite{wang2024learnable} to align them with \gls{llm}s and the objectives of recommendation, as well as to preserve interpretability~\cite{fang2025hid}. On the other hand, textual representations are not only fully interpretable but also editable and easy to adapt with minimal levels of finetuning (see Table~\ref{table:methods}). Further, textual embeddings don't require extensive version control unlike embeddings where different versions may be incompatible with each or require auxiliary objectives/parameters to make them compatible.

Conventionally, member representations are realized as either dense, learned vectors~\cite{zhao2023embedding} or high-dimensional sparse features~\cite{dash1997feature}. Dense representations are produced by parametric encoders $f_{\theta} : \mathcal{X}\!\to\!\mathbb{R}^{d}$, such as contrastive or supervised embedding networks, and then consumed by downstream predictors $g_{\phi}$ or nearest neighbor indices~\cite{huang2020embedding}. Two fundamental limitations emerge from this design. First, the latent dimensions of $f_{\theta}(X)$ are generally not semantically grounded, which impedes interpretability and hinders manual inspection. Second, modifications to the encoder backbone ($\theta \to \theta'$) induce a distributional shift in the embedding space ($\mathbb{P}_{f_{\theta}(X)}\!\to\!\mathbb{P}_{f_{\theta'}(X)}$), necessitating repeated downstream recalibration or full retraining of $g_{\phi}$, and incurring substantial operational costs due to re-embedding corpora and rebuilding indices~\cite{liu2025scalable}.
%(an $O(N)$ process for datasets of size $N$) \cite{liu2025scalable}. 
In contrast, sparse, hand-engineered features mitigate opacity but suffer from brittle feature coverage and high maintenance overhead.

\begin{table}
\centering
\caption{Comparing common user representation methods.}
\vspace{-6pt}
\label{table:methods}
\scalebox{0.88}{%
\begin{tabular}{lcc} 
\toprule
\textbf{Method} & \textbf{Interpretable} & \textbf{Plug \& Play} \\
\midrule
Embeddings & \ding{55} & \ding{55} \\
Semantic ID & \partialmark & \ding{55} \\
Text &\ding{51} & \ding{51} \\
\bottomrule
\end{tabular}
}
\vspace{-2pt}
\end{table} 

The recent emergence of \gls{llm}-centric retrieval and ranking pipelines further exacerbates these incompatibilities. Token-based LLMs operate natively in text space, so integrating vectorized or sparse representations often requires intermediary projection layers, auxiliary alignment objectives, or \gls{llm} finetuning, each of which increases latency, parameters, and coupling across components. Semantic IDs~\cite{singh2024better} partially bridge this gap by providing more interpretable, token-like identifiers, but empirical work shows they still require substantial finetuning and alignment to preserve interpretability and task performance~\cite{wang2024learnable,fang2025hid}.

By contrast, compact textual representations reside directly in the \gls{llm} token space: they are human-interpretable, editable, and forward-compatible with prompt-based and retrieval-augmented \gls{llm} usage patterns, thereby substantially reducing the burden of version management and cross-component alignment (see Table~\ref{table:methods} for a comparative summary). This transition from opaque vector spaces to concise textual representations necessitates principled selection mechanisms, such as RL guided by engagement-driven rewards, to ensure brevity without loss of predictive signal.

% \begin{table}[t!]
%   \caption{Comparing Common User Representation methods}
%   \label{tab.case_study}
%   \centering
%   \begin{tabular}{|l|c|c|}
%     \hline
%     Method & Interpretable & Plug \& Play \\
%     \hline
%     Embeddings  & \textcolor{red}{\ding{55}} & \textcolor{red}{\ding{55}} \\
%     Semantic ID & \partialmark                 & \textcolor{red}{\ding{55}} \\
%     Text        & \textcolor{green}{\ding{51}} & \textcolor{green}{\ding{51}} \\
%     \hline
%   \end{tabular}
% \end{table}

% \begin{table}
% \centering
% \begin{tabular}{|c||c|c|}
% \hline
% \textbf{Method} & \textbf{Interpretable} & \textbf{Plug \& Play} \\ 
% \hline
% Embeddings & \ding{55} & \ding{55} \\ 
% \hline
% Semantic ID & \partialmark & \ding{55} \\
% \hline
% Text &\ding{51} & \ding{51} \\
% \hline
% \end{tabular}
% \caption{Comparing common user representation methods.}
% \label{table:methods}
% \end{table}

\begin{comment}
\begin{table}
\caption{Comparing common user representation methods.}
\vspace{-8pt}
\small{
\centering
\begin{tcolorbox}[enhanced, colback=white, colframe=white, width=\linewidth, boxrule=0pt, top=0pt, bottom=0pt]
\centering
\renewcommand{\arraystretch}{1.1}
\begin{tabular}{ccc}
\hline
\textbf{Method} & \textbf{Interpretable} & \textbf{Plug \& Play} \\
\hline
Embeddings & \ding{55} & \ding{55} \\
\hline
Semantic ID & \partialmark & \ding{55} \\
\hline
Text &\ding{51} & \ding{51} \\
\hline
\end{tabular}
\end{tcolorbox}
}
\vspace{-8pt}
\label{table:methods}
\end{table} 
\end{comment}

Although the context windows of modern \gls{llm}s have grown substantially, naively concatenating all available member text is neither computationally nor algorithmically tenable. From a systems perspective, attending to unbounded token sequences increases serving costs and violates latency constraints. From an algorithmic perspective, long, redundant contexts dilute discriminative signals. Concatenating multiple sources often introduces redundant or low-utility content, which degrades nearest-neighbor retrieval precision and weakens downstream ranking models by increasing input noise and reducing the effective signal-to-noise ratio.

We therefore cast the problem as a text-generation task: given a set of heterogeneous textual sources for a member, we learn a stochastic policy to produce a \textit{compact textual synopsis} that maximizes expected downstream utility. As noted above, information across multiple sources often exhibits redundancy. For example, many sections in web profiles and professional sources may be identical. User activities also tend to indicate broad intents that can be captured precisely. Our approach relies on the intuition that an effective user representation capturing all salient information across multiple sources should be sufficient to reliably predict the user's future job-related actions. Uniquely, we avoid external supervision, relying solely on engagement labels to guide generation. We evaluate these representations qualitatively using an \gls{llm}-as-a-judge framework and quantitatively via downstream ranking performance. 

\section{Related Works}
\subsection{LLMs for RecSys}
\gls{llm}s are trained on vast amounts of data and thus implicitly capture world knowledge. Unsurprisingly, they have emerged as a cornerstone of modern \gls{recsys} in recent years. A number of works~\cite{ren2024representation,liu2025scalable,10.1145/3711896.3737029} use \gls{llm}s to create dense representations that are used by downstream models. Beyond representation learning, \gls{llm}s have also been employed as retrievers~\cite{lee2025gemini}, rankers~\cite{luo2025recranker, ye2025applying}, or both~\cite{tang2024self, shah2025towards}. In addition, they have been applied to data labeling and quality evaluation \cite{10.1145/3690624.3709413}. LLMs are also enabling a new paradigm of agentic recsys \cite{huang2025towards, palumbo2025you}.  Li et al.~\cite{li2023text} demonstrated that sequential recommendation can be entirely modeled in textual form. This finding has inspired a growing body of work adapting \gls{llm}s to generative recommenders, where items are directly generated rather than retrieved or ranked~\cite{ji2024genrec}. With their increasing adoption, it is prudent to develop techniques that facilitate integration and rapid iteration, thereby aiding the practical deployment of \gls{llm}-based recommenders.

\subsection{User Representations}
\subsubsection{Embeddings} Embeddings have been the de facto modality for user representation across industries~\cite{qiu2021u, wang2022learning, zhang2023twhin}. Li et al.~\cite{li2024calrec} use a two-stage finetuning strategy combining language modeling and contrastive losses to generate dense user embeddings. Similarly, Zhang et al.~\cite{zhang2024embsum} leverage a pretrained Mixtral model to summarize user action histories, which are then used as an auxiliary objective for embedding generation via a T5 backbone. Liu et al.~\cite{liu2025scalable} utilize \gls{llm}-based embeddings as node features in a GNN to obtain holistic representations across multiple data sources. DV365 \cite{lyu2025dv365} and Ju et al.~\cite{ju2025learning}  propose a unified embedding that aggregates information from multiple domains.

\subsubsection{ID-based methods}
%\textbf{ID-based methods.}
A number of works utilize member IDs directly to enable personalization in downstream models. For example, P5~\cite{geng2022recommendation} trains a foundation model in a T5-like text-to-text format, where users are represented by their IDs acting as tokens. Singh et al.~\cite{singh2024better} introduce Semantic IDs, in which item codebook IDs from an RQ-VAE are used to model users.

\subsubsection{Textual Representation}
%\textbf{Textual Representation.}
Recent works have explored representing users directly in text form. LACE~\cite{mysore2023editable} extracts readable concepts via retrieval from user interactions. Tears~\cite{penaloza2024tears} learns user activity summaries using an optimal transport procedure with respect to VAE embeddings. Gao et al.~\cite{gao2024end} employ an offline \gls{rl} algorithm, similar to DPO, to infer textual profiles from user activities. RecGPT~\cite{zhang2024recgpt} adopts language-modeling-style pretraining to instill user ID knowledge into its architecture. 

\subsection{RL for RecSys}
\gls{rl} has been applied to sequential recommendation problems even before the advent of \gls{llm}s~\cite{shani2005mdp, moling2012optimal, afsar2022reinforcement, taghipour2007usage}. Several works have focused on state space representation. Following the introduction of DeepSeek-R1~\cite{guo2025deepseek}, \gls{rl} was employed to augment reasoning in ranking systems~\cite{zhuang2025rankr1enhancingreasoningllmbased}. Rec-R1~\cite{lin2025rec} proposes a general \gls{rl} framework to enhance input queries and directly optimize downstream performance. Wang et al.~\cite{wang2024reinforcement} model user actions and states with \gls{rl}, thereby reducing the need for large amounts of training data. Rank-R1~\cite{zhuang2025rankr1enhancingreasoningllmbased} leverages \gls{rl} to utilize knowledge graphs more effectively. Mao et al.~\cite{mao2025reinforced} also model users with \gls{rl}, but formulate the problem as prompt editing to augment \gls{llm} responses with user-specific hints. Finally, Wang et al.~\cite{wang2025policy} leverage \gls{rl} to isolate causally relevant features, which are then used to refine user embeddings.

\begin{figure*}
\centering
\includegraphics[width=0.95\linewidth,height=5.8cm]{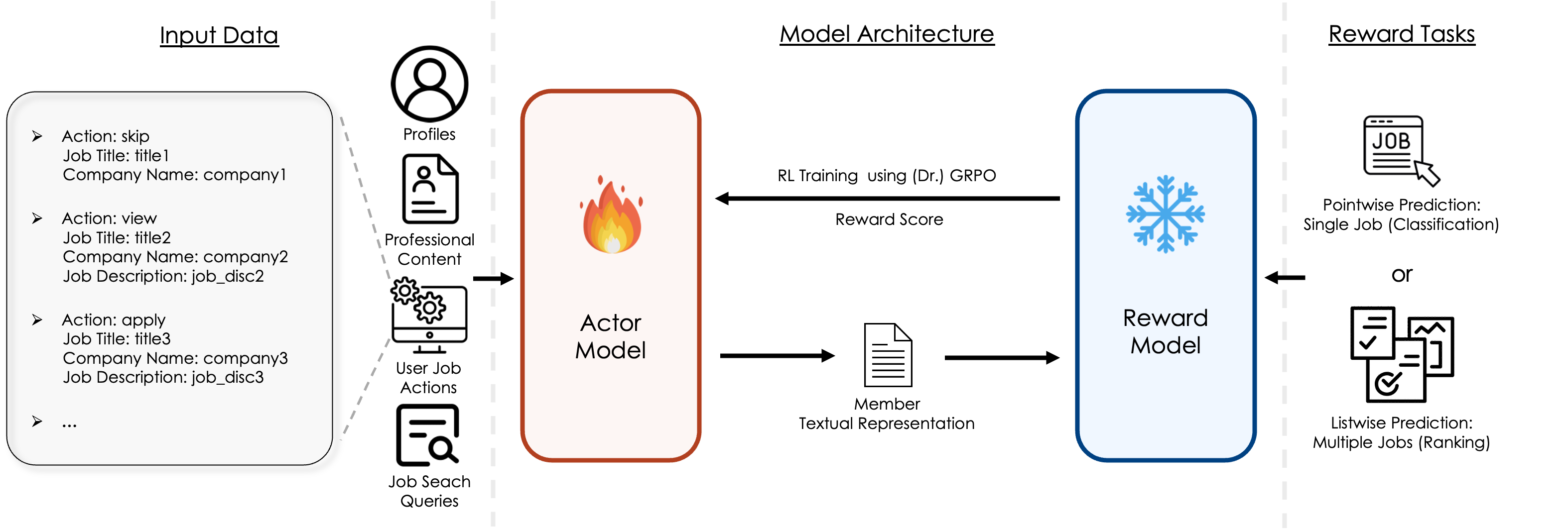}
\caption{\gls{rl} framework for member textual representation generation. The actor, a pre-trained \gls{llm} finetuned via reinforcement learning, generates textual representations from member data, while the frozen reward model scores them based on their alignment with new job posting(s) (either pointwise or listwise prediction). The resulting reward signal is used to optimize the actor. When a member skips a recommended job, the job description is excluded to reduce input token length.}
\label{fig:workflow}
\end{figure*}

\section{Proposed Methodology}
\subsection{Preliminaries}

\gls{rl} provides a general framework for sequential decision-making~\cite{kaelbling1996reinforcement,sutton1998reinforcement}.
Model-free methods directly optimize policies through interaction with the environment and include
value-based algorithms such as Q-learning~\cite{watkins1992q,mnih2013playing} as well as policy
optimization methods~\cite{williams1992simple,sutton1999policy}.
In this work, we focus on model-free policy optimization, which has proven particularly effective
for aligning \gls{llm}s with task-specific objectives and human preferences.

In the context of \gls{llm}s, the policy $\pi_\theta$ defines a conditional distribution over tokens. The optimization objective is to maximize the expected return:
\[
\mathcal{J}(\theta)=\mathbb{E}_{\tau\sim\pi_\theta}[R(\tau)],
\quad
\nabla_\theta \mathcal{J}(\theta)\approx
\mathbb{E}\!\left[\sum_t \nabla_\theta \log \pi_\theta(a_t\mid s_t) A_t \right],
\]
where $A_t$ denotes an advantage estimator used for variance reduction~\cite{schulman2015trust}.
For \gls{llm}s, the policy defines a conditional distribution over tokens, actions correspond to
token selections, and trajectories correspond to generated sequences.

In long-horizon autoregressive generation, naive application of policy gradient methods~\cite{williams1992simple}
can result in high-variance updates, motivating stabilized objectives such as
\gls{ppo}~\cite{schulman2017proximal}.
\gls{ppo} constrains policy updates via a clipped surrogate objective and KL regularization,
balancing stability and sample efficiency, but requires a learned critic, which can be costly at scale.

\gls{grpo}~\cite{shao2024deepseekmath} simplifies training by eliminating the critic and instead
using group-relative baselines.
For each query $q$, a group of candidate outputs $\{o_1,\dots,o_G\}$ is sampled from the old policy
$\pi_{\theta_{\text{old}}}$ and scored by a reward function.
The group-average reward serves as a baseline, yielding token-level advantages.
Each token $o_{i,t}$ is treated as an action conditioned on the prefix $(q,o_{i,<t})$, and the
policy is updated by maximizing the following PPO-style objective:
\begin{comment}
\begin{equation*}
\scalebox{0.78}{$
\begin{split}
\mathcal{J}_{\text{GRPO}}(\theta)
&=
\mathbb{E}_{q \sim \mathbb{P}(Q), \{o_i\}_{i=1}^G \sim \pi_{\theta_{\text{old}}}(O\mid q)}
\Bigg[
\frac{1}{G}\sum_{i=1}^G \frac{1}{|o_i|}\sum_t
\min\!\Big(
r_{i,t}\hat{A}_{i,t},
\text{clip}(r_{i,t},1-\epsilon,1+\epsilon)\hat{A}_{i,t}
\Big)
\\
&\qquad
-\beta \mathbb{D}_{\mathrm{KL}}(\pi_\theta\Vert\pi_{\text{ref}})
\Bigg],
\end{split}
$}
\end{equation*}
\end{comment}
\begin{equation*}
\scalebox{0.78}{$
\begin{split}
\mathcal{J}_{\text{GRPO}}(\theta)
&=
\mathbb{E}_{q \sim \mathbb{P}(Q), \{o_i\}_{i=1}^G \sim \pi_{\theta_{\text{old}}}(O\mid q)}
\\
&\quad
\Bigg[
\frac{1}{G}\sum_{i=1}^G \frac{1}{|o_i|}\sum_t
\min\!\Big(
r_{i,t}\hat{A}_{i,t},
\text{clip}(r_{i,t},1-\epsilon,1+\epsilon)\hat{A}_{i,t}
\Big) -\beta \mathbb{D}_{\mathrm{KL}}(\pi_\theta\Vert\pi_{\text{ref}})
\Bigg],
\end{split}
$}
\end{equation*}
where $r_{i,t} =
\pi_\theta(o_{i,t}\mid q,o_{i,<t}) /
\pi_{\theta_{\text{old}}}(o_{i,t}\mid q,o_{i,<t})$,
$\epsilon$ controls the clipping range, and $\hat{A}_{i,t}$ denotes the group-relative token-level
advantage~\cite{shao2024deepseekmath}. 

We further integrate recent advances in stabilizing RL-based sequence optimization, including \gls{dapo}~\cite{yu2025dapo}, which applies asymmetric clipping to prevent premature entropy collapse and preserve exploration, and Dr.~\gls{grpo}~\cite{liu2025understanding}, which explicitly compensates for systematic length bias in the policy gradient objective. This combination yields a favorable trade-off between  the optimization stability benefits of PPO-style updates and the computational efficiency required for scalable large-vocabulary \gls{llm} decoding.

\subsection{\gls{rl} Setup}
We formalize the generation of user representations as a text-generation problem trained via reinforcement learning, building on training paradigms such as Rec-R1~\cite{lin2025rec}. Each context $q$ aggregates heterogeneous, potentially variable-length textual sources that together characterize a member's job-seeking behavior:
\begin{align*}
q \;=\; \{\, & \text{profile attributes},\ 
\text{professional content},\ \\
& \text{job search actions},\ 
\text{search queries in jobs page}\,\}.
\end{align*}
A detailed breakdown of these data sources is provided in Section~\ref{subsec:data}. Given $q$, we sample a textual synopsis $o\in\mathcal{O}$ from a policy $\pi_\theta(\cdot\mid q)$, where $\mathcal{O}$ denotes the space of admissible textual representations. Our desiderata for $o$ are following: (i) \textbf{salience}: it must retain all of the information predictive of future job interactions, and (ii) \textbf{conciseness}: it must pertain to a strict token budget, ensuring efficiency under production latency constraints. To put it in words, the ideal summary should have all the necessary information to predict user action on unseen jobs, and yet be succinct. We operationalize these desiderata via a scalar reward function $R(o\mid q)$ and optimize the expected reward:
$$
\max_{\theta}\; \mathbb{E}_{q\sim\mathbb{P}(Q)}\; \mathbb{E}_{o\sim\pi_\theta(\cdot\mid q)}\big[\,R(o\mid q)\,\big].
$$
The end-to-end pipeline is summarized in Figure~\ref{fig:workflow}. Concretely, the policy $\pi_{\theta}$ is implemented as an LLM actor and finetuned with model-free policy optimization. To enable online deployment, the actor is a medium-capacity 1.7B parameter language model, conditioned via an instruction-style prompt that elicits a compact synthesis of job-relevant signals from the multi-source user context. 
% For confidentiality we only describe the prompt's goals at a high level.
Rather than relying on expensive and scarce human labels, we employ a \textit{labeling-free} supervision strategy. 
Generated synopses $o$ are scored by a promptable high-capacity LLM evaluator that serves as a preference oracle, mapping $(s,o)\mapsto R(s,o)$, a scalar reward reflecting the quality of $o$ with respect to the job seeker’s contextual state $s$.
The reward is designed to align with downstream signals of user preference -- such as relative ranking and observed interaction outcomes -- described in Section \ref{subsec:reward_design}, and serves as the primary supervision signal for RL training. The reward prompt and a worked example are provided in Appendix~\ref{appendix:prompt_templates}.

\begin{figure}[t]
    \centering
    \includegraphics[width=0.38\textwidth]{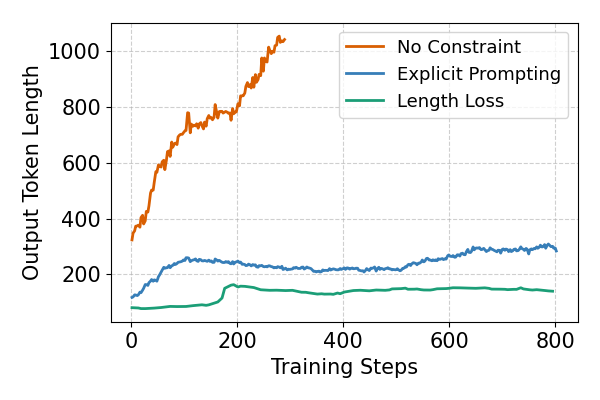}
\vspace{-3pt}
    \caption{Output token lengths are compared under three conditions: no constraint, explicit prompting, and using a length loss with a quadratic penalty function. Without any constraint, the output length grows rapidly, and we therefore stop the training process early. Explicit prompting substantially reduces this growth, although a slight upward trend persists and precise control over output length is not achieved. Applying the length penalty function stabilizes output length around $150$ tokens.}
    \label{fig:training_length}
\vspace{-3pt}
\end{figure}

Choosing the reward oracle involves an explicit engineering trade-off. One option is to SFT smaller models on task-specific data: this reduces per-step RL compute and inference latency but requires collecting and maintaining supervised labels for each downstream objective. The alternative is to use a large, generalist LLM that provides strong zero-/few-shot semantic judgments across tasks, removing the need for per-task SFT at the cost of increased evaluation compute. We adopt the latter strategy and instantiate the oracle with a larger model with 30 Billion active parameters. To verify it's domain-specific reliability, we benchmark it against GPT-o1 with Chain-of-Thought~\cite{wei2022chain} since o1 is well recognized for its multi-step reasoning and semantic understanding. Results on both pointwise and listwise reward tasks (see Section~\ref{subsec:reward_design}) show comparable performance and good alignment with logged engagement metrics.

%Practically, using a large oracle increases RL evaluation cost; this overhead can be amortized through batched scoring, or selective re-evaluation.
Crucially, treating the reward model as a promptable preference oracle provides a flexible, label-light supervision mechanism that simplifies deployment across evolving downstream tasks while preserving a semantically grounded supervision signal for policy optimization.

\begin{figure*}[htbp]
    \centering
    \begin{subfigure}[b]{0.32\textwidth}
        \includegraphics[width=\textwidth]{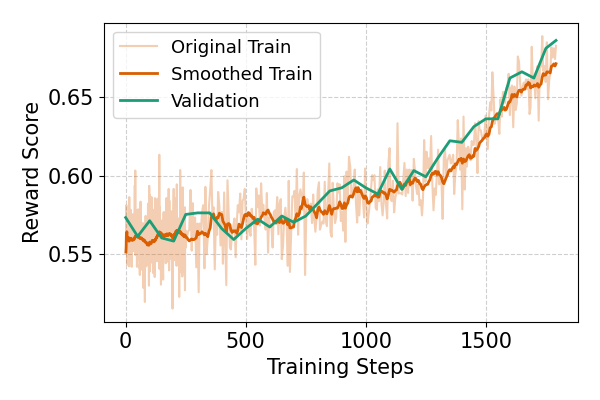}
        \caption{Pointwise string-based reward}
        \label{fig:poinwise_reward_binary}
    \end{subfigure}
    \begin{subfigure}[b]{0.32\textwidth}
        \includegraphics[width=\textwidth]{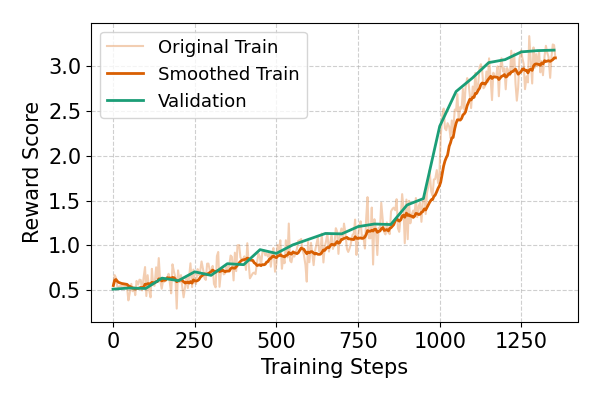}
        \caption{Pointwise logprob-based reward}
        \label{fig:poinwise_reward_logprob}
    \end{subfigure}
    \begin{subfigure}[b]{0.32\textwidth}
        \includegraphics[width=\textwidth]{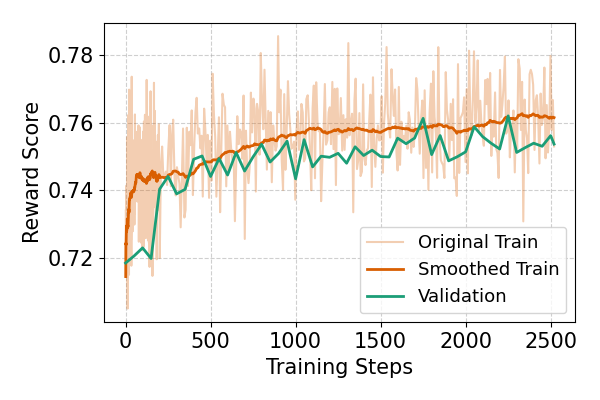}
        \caption{Listwise reward}
        \label{fig:listwise_reward_ndcg}
    \end{subfigure}
    
    \begin{subfigure}[b]{0.32\textwidth}
        \includegraphics[width=\textwidth]{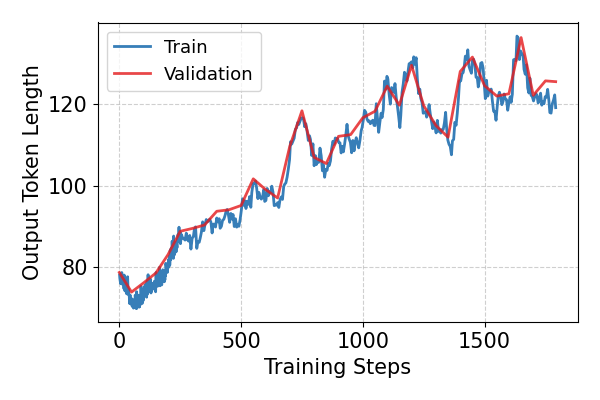}
        \caption{Pointwise string-based length}
        \label{fig:poinwise_length_binary}
    \end{subfigure}
    \begin{subfigure}[b]{0.32\textwidth}
        \includegraphics[width=\textwidth]{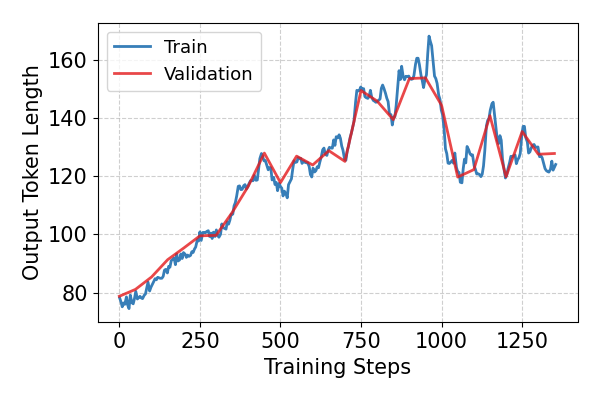}
        \caption{Pointwise logprob-based length}
        \label{fig:poinwise_length_logprob}
    \end{subfigure}
    \begin{subfigure}[b]{0.32\textwidth}
        \includegraphics[width=\textwidth]{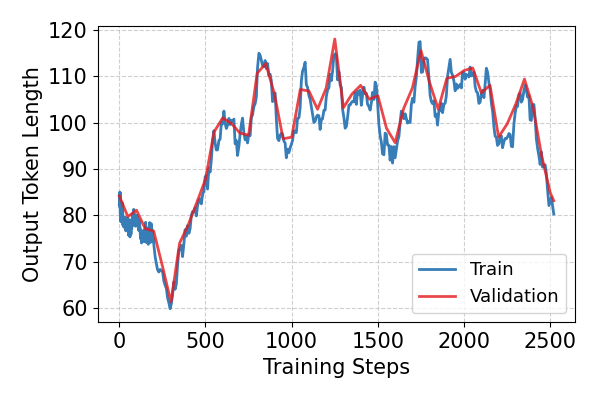}
        \caption{Listwise length}
        \label{fig:listwise_length_ndcg}
    \end{subfigure}
    \caption{Evolution of the reward score and the corresponding output token length on the training and validation datasets for pointwise and listwise tasks. The training curve is smoothed using a moving average with a window size of 50.}
    \label{fig:reward}
\end{figure*}
\subsection {Reward Design}
\label{subsec:reward_design}
In this section, we introduce the reward design adopted in our framework. Reward models are used to obtain scores that are treated as proxies for the desired objective. Motivated by classical optimization objectives in recommendation systems, we extend commonly used pointwise and listwise prediction tasks to reward modeling as well and describe each below.

\subsubsection{Pointwise Reward}
In the pointwise setting, we assess the model’s ability to predict whether a user will take a specific action on an individual item. This is cast as a binary classification problem: given a user’s summarized textual representation alone, predict whether they will apply to a target job (see Table~\ref{table:reward_class_template}). We introduce two reward formulations to guide policy learning.

\textbf{String-based reward}. It provides a sparse signal based solely on the model’s final prediction $\hat{y}$. The reward is defined as
\[
\hspace{3em}
R = \mathds{1}\{\hat{y} = y\},
\] 
where $y \in \{0,1\}$ denotes the ground-truth label. This formulation rewards exact correctness but offers no gradient of confidence.

\textbf{Log-probability rewards}. This incorporates the model’s predictive confidence to provide a denser and more informative signal. Let $\log p(\text{yes})$ and $\log p(\text{no})$ denote the log-probabilities of positive and negative outcomes. The reward is given by
\[
\hspace{3em}
R = (2y - 1) \big(\log p(\text{yes}) - \log p(\text{no})\big), \quad y \in \{0,1\}.
\] 

This formulation yields a continuous confidence-aligned reward, in contrast to the binary outcome of the string-based approach.
In practice, inference efficiency poses constraints. For example, when using frameworks such as vLLM~\cite{kwon2023efficient}, low-rank tokens may be truncated from the returned distribution. In such cases, we assign a log-probability of $-\infty$ to missing tokens. To ensure stable training, we clip rewards to the interval $[-4,6]$. These bounds were determined empirically to prevent extreme values from dominating the policy updates.

%The evolution of the reward score during training for the two approaches is illustrated in Figures~\ref{fig:poinwise_reward_binary} and~\ref{fig:poinwise_reward_logprob}.

\subsubsection {Entropy-Weighted Listwise Reward}
While pointwise rewards capture predictive accuracy at the level of individual interactions, they do not model relative item preferences.
In contrast to the pointwise reward, the listwise reward is defined at the level of a set of items. In a listwise prediction task, given several new job postings for a member, the model outputs a ranking of these jobs according to the likelihood of application (see Table~\ref{table:reward_rank_template}). In this setting, listwise labels distinguish among three types of actions: apply, view, and skip, whereas pointwise labels only indicate whether a member applies to a job, treating view and skip as not apply. The listwise reward is defined as the \gls{ndcg} of the predicted ranking, with relevance scores assigned as 5 for apply, 1 for view, and 0 for skip. This reward reflects how well the predicted order aligns with the ground truth relevance.

However, not all training samples are equally discriminative. A list containing all ``Skips'' offers less supervision signal than a diverse list of ``Applies'' and ``Views''. To address this, we introduce a novel \textit{Entropy Weighting} mechanism. In order to provide stronger reward signals, we weight the listwise reward by the entropy of the label distribution within each list. Specifically, let the possible labels for each job be $L = \{\text{apply}, \text{view}, \text{skip}\}$, and let $f_l$ denote the fraction of jobs in the list with the label \(l \in L\), that is, $f_l = \frac{\text{\# of jobs with label } l}{n}$, where $n$ is the length of the list. The Shannon entropy of the list is then computed as
\[
H = - \sum_{l \in L} f_l \ln f_l.
\]
Lists with more diverse labels, and thus higher entropy, are considered more informative and are assigned greater weight in the overall reward. Using entropy to weight the rewards empirically speeds up the convergence significantly. %For example, the sequence $\{\text{apply}, \text{skip}, \text{apply}, \text{view}, \text{skip}\}$ carries more weight than $\{\text{skip}, \text{skip}, \text{skip}, \text{skip}, \text{view}\}$, reflecting its higher information content. 
The final weighted listwise reward for a job list is computed as
\[
R = H \cdot \text{\gls{ndcg}}.
\]
%Figure~\ref{fig:listwise_reward_ndcg} depicts the evolution of the weighted \gls{ndcg} during training.

\subsubsection {Constraints \& Regularization}
\label{subsubsec:length_control}
One challenge in reinforcement learning with language models is policy degeneration, where the policy exploits imperfections in the reward signal. In our setting, this appeared as uncontrolled verbosity: the actor discovered that it could inflate the expected reward from the evaluator by generating excessively long outputs (please refer to Figure~\ref{fig:training_length}). This constitutes a form of reward hacking ~\cite{gao2023scaling, ziegler2019fine}, as longer sequences are more likely to include reward-relevant tokens.
%\textcolor{red}{Should we include the output summaries of these three in the appendix for comparison?} \textcolor{blue} {let's have summary samples from final models in the actual paper after disambiguation, can use placeholder for now, if we run out of space we move to appendix but it is better to have them in main paper)}.%
% This typically occurs when the reward model struggles to evaluate the generated representations for the given task, causing the actor model to over-generate text. 
To mitigate this issue, we employ strategies that build on Dr. \gls{grpo}~\cite{liu2025understanding}, which stabilizes policy updates and prevents uncontrolled growth in output length. To this end, we implement two complementary approaches to further regulate output length.

    \textbf{Explicit Prompting}. We modify the actor prompt to explicitly request concise outputs, such as ``A concise 2 to 3 sentence summary that serves as a member representation''; this conditions the policy's prior distribution and narrows the generative search space to favor compact outputs.
    
    \textbf{Length Loss}. Inspired by reward shaping in reinforcement learning~\cite{ng1999policy}, we augment the base reward with a length penalty $f_{\text{length}}(|o|)$, where $|o|$ denotes the number of tokens in the output. For example, $f_{\text{length}}(|o|) = -(|o|-150)^2 , |o| > 150$. No penalty is applied when the output length is below 150 tokens, and the penalty increases quadratically as the output length grows beyond 150. %An illustration of the penalty function is shown in \textcolor{red}{Figure~\ref{fig:penalty_func}}. 
    The total reward is defined as $R_{\text{total}} = R + \lambda \cdot f_{\text{length}}(|o|)$, where $\lambda$ is a tunable hyperparameter, and we set $\lambda = 10^{-5}$ in our experiments. This discourages excessively long outputs while maintaining informativeness.

As shown in Figure~\ref{fig:reward}, this approach effectively mitigates length explosion and stabilizes finetuning across both pointwise and listwise objectives. By directly addressing verbosity, our method extends prior efforts in RLHF regularization and provides a practical mechanism for producing compact, informative, and deployable user representations in web-scale systems.

\begin{table*}[h!]
\caption{
Policy drift under different reward formulations in our RL approach. Input data is synthetically generated by replacing real entities with surrogates and do not correspond to any real user. Newly introduced information in the generated summaries is shown in \textbf{bold}. 
(i) The \emph{Naive summary} relies almost exclusively on profile information, lacking awareness of which sources are most useful for downstream tasks.
(ii) The \emph{Pointwise string-based reward} encourages summaries that better capture user skills, while also reflecting activity trends and listing representative companies and titles under consideration. 
(iii) The \emph{Pointwise Logprob-based reward} exhibits signs of reward hacking, with activity trends overemphasized at the cost of other input sources. 
(iv) The \emph{Listwise reward} produces more conservative summaries than pointwise rewards, but still captures a richer set of skills and companies the user has shown interest in.
Text inside <...> tags has been redacted or replaced to prevent accidental attribution.
% \textcolor{red}{Option 2. From three opetions, I like this one the most. but option 3 is also good. which one do you prefer.}\textcolor{blue}{i agree! lets go with this template}
}
\vspace{-8pt}
\centering
%\footnotesize{
\small{
\begin{tcolorbox}[enhanced, width=\linewidth]
\textbf{Vanilla:} \\[0.3em]
A skilled System Administrator with 3.5 years of experience in cloud and on-premise infrastructure, proficient in Linux, Azure, and network administration, targeting roles in cloud engineering, database administration, and IT support with a focus on security and automation.

\bigskip\hrule\bigskip

\textbf{Pointwise String-based Reward:} \\[0.3em]
The member is a System Administrator with experience in \textbf{Windows, Linux, and Azure cloud environments, specializing in server management, security, and automation}. They are actively applying for \textbf{multiple cloud and IT support roles, including Cloud Engineer, Technical Support Engineer, and Site Reliability Engineer}. The member is highly skilled in Linux system administration, Azure, and network security, \textbf{with a focus on monitoring, troubleshooting, and optimizing infrastructure}. They are targeting roles that require cloud expertise, database administration, and SRE experience, with a \textbf{strong interest in companies like \textless{}company\_1\textgreater{}, \textless{}company\_2\textgreater{}, and \textless{}company\_3\textgreater{}}. \textbf{The member is persistent in applying to various positions but has viewed some job listings, indicating a cautious approach}.

\bigskip\hrule\bigskip

\textbf{Pointwise Logprob-based Reward:} \\[0.3em]
The member has a clear and consistent interest in roles related to system administration, cloud engineering, and database administration. \textbf{They have applied to multiple positions in these areas, including Cloud Engineer, Database Administrator, and Support Engineer. Their profile, experience, and historical activities all point to a focus on IT infrastructure, cloud platforms, and database management. The member has a strong track record of applying to jobs in these categories and has a history of applying to similar roles}. Given the consistent interest and application behavior, the member is highly relevant to any job description in the cloud engineering, database administration, or system administration categories. The relevance score for these job types is extremely high, making them the top priority for the member

%"The member is a Full Stack Software Developer with 4 years of experience at Sharaan Infosystems, Relex Health, and Agiledge Solutions. They are actively applying to various Angular Developer, NodeJS Developer, and Full Stack Engineer positions, with a focus on companies in Bengaluru. The member is highly skilled in Angular, Node.js, and SQL, and has a strong track record of performance optimization and API integration. They are seeking a mid-level Angular Developer role at Uplers, but have applied to multiple similar positions, indicating a broad job search with a focus on frontend and backend development."

\bigskip\hrule\bigskip

\textbf{Listwise Reward:} \\[0.3em]
A System Administrator seeking a Cloud/Support Engineer role in \textbf{\textless{}user\_location\textgreater{}}, with focus on Database Administration and Azure cloud services. Candidate has experience in \textbf{Linux/Windows server administration, network management, and cloud provisioning. Targeting positions at companies like \textless{}company\_1\textgreater{}, \textless{}company\_2\textgreater{}, and \textless{}company\_4\textgreater{}, with a strong background in  Cloud Infrastructure and Azure}. Seeking a role that aligns with their skills in cloud ETL, database administration, and system support.

% "The member is seeking a Mid Angular Developer role at a company in Bengaluru, Karnataka, India. They are looking for a Full Stack Engineer with 3+ years of experience, specializing in Angular, Node.js, and SQL databases. The candidate prefers companies in the tech sector, with a focus on ERP applications, and is open to immediate joiners. They are actively applying to positions like Angular Developer, NodeJS Developer, and Full Stack Engineer."

\end{tcolorbox}
}
\vspace{-6pt}
\label{table:reward_box_example}
\end{table*}

% \begin{table}[t]
% \centering
% \begin{tcolorbox}[myblock,width=0.95\columnwidth,colback=gray!4]
% \footnotesize
% The member is seeking a Warehouse Manager role with experience in quality assurance, team leadership, and safety management.
% They have recently viewed a \redactbar{SkillBridge} job posting for a Warehouse Manager at \redactbar{US Foods} and are actively
% searching for terms like ``manager,'' ``assistant,'' and ``Sales Consultant.'' Their profile highlights a background in logistics,
% safety, and customer service, with recent experience at \redactbar{JEAR Logistics} and \redactbar{Coca-Cola Consolidated}.
% They are transitioning from contractor to full-time roles and have skills in lead generation, social media management, and rapid learning.

% \medskip
% \textbf{Relevance Ranking:}\\
% 1.~Warehouse Manager\\
% 2.~Assistant Manager\\
% 3.~Sales Consultant\\
% 4.~Manager (general)\\
% 5.~Assistant Manager
% \end{tcolorbox}
% \caption{Emergence of Relevance Ranking when text is produced without constraints leading to feature discovery.  Portions of the text have been redacted for privacy, indicated by black boxes.}
% \label{table:redacted_example}
% \end{table}

\subsubsection {Format Penalty}
\label{subsubsect:format_penalty}
We notice that the actor model sometimes adds auxiliary information to guide the reward model towards making the correct decision, for example, a predicted list of jobs a user may like (see Section~\ref{sec:emergent_behavior}), which may not always be strongly grounded in user input. In our experiments, this auxiliary information is consistently presented as a new paragraph in the generated text. Thus, we assign a reward of $-1$ whenever the actor produces multiple paragraphs. We find that this simple constraint is sufficient to ensure that the model's output remains strongly grounded.

\subsection{Dataset}
\label{subsec:data}
We construct our dataset by subsampling one month of multi-faceted user data and job actions on the platform similar to \cite{liu2025scalable}. To ensure a strict temporal separation, we form the validation set by sampling from the subsequent two weeks of logs. Job engagements include user actions on jobs, namely apply, view, or skip. This engagement history is used to construct both historical activities as model inputs as well as labels for the reward task. 
The data we use is managed in accordance with relevant member settings to ensure full respect for their privacy choices.

In the pointwise setting, the action of the most recent activity serves as the ground truth, and the information of the engaged job in that activity, including job title, company name, and job description, is provided as input to the reward model. We downsample the training data to maintain 1:1 ratio between applies and non-applies.

In the listwise setting, information from the last five activities is used as input to the reward model, with the corresponding actions serving as the label. Similar to the pointwise setting job descriptions, job title, and job company names and are provided as inputs to reward model. We remove any trivial samples in which all the jobs to be ranked have the same label.

We use roughly 200k samples for training. On average, input for each member spans ~5000 tokens over all sources. All features are joined in a time-aware manner, i.e., we use feature values at the time of action to not pollute our metrics due to time-travel.

\subsection{Training Details}
We implement and evaluate our method using the \href{https://github.com/volcengine/verl}{VeRL} framework~\cite{sheng2025hybridflow}, a flexible \gls{rl} training library designed for large language models.
Policy optimization is performed using Dr. \gls{grpo}\cite{liu2025understanding}.

To maximize throughput and reduce parameter offloading overhead, we distribute computation across a two-node cluster with 16 H200 GPUs. Actor optimization, rollout generation, and control logic run on 8 GPUs, while the larger reward model occupies the remaining 8 GPUs. This separation decouples expensive reward evaluation from actor forward and backward passes, improving overall throughput. For reward-model inference we use vLLM~\cite{kwon2023efficient}, which enables faster batched scoring.

Each prompt is expanded into 4 sampled responses to enable group-based evaluation under Dr. \gls{grpo}\cite{liu2025understanding}. The policy is trained for 8 epochs with a learning rate of $3\times10^{-7}$, a global mini-batch size of $768$, a micro-batch size of $32$, and a KL regularization factor of 0.001 to constrain divergence from the pretrained prior. 
During early training we observed entropy collapse, which we countered using the \gls{dapo}~\cite{yu2025dapo} variant with $\epsilon_{\text{high}}=0.28$; its asymmetric clipping stabilizes updates and preserves exploration by preventing premature loss of output diversity.

\begin{table}[t]
\centering
\caption{Feature discovery from generated text.}
\vspace{-12pt}
\caption*{\footnotesize Emergence of a ranked sample generated by unconstrained RL. The text was produced without explicit prompting, suggesting that the resulting ranked list may serve as a potentially useful feature. All entities were replaced with synthetic surrogates, denoted by [ENTITY\_TYPE] prior to inference.}
\label{table:redacted_example}
\vspace{-12pt}
\begin{tcolorbox}[myblock,width=\columnwidth,colback=gray!4,top=2.5pt,bottom=-5pt]
\footnotesize
The member is seeking a Warehouse Manager role with experience in quality assurance, team leadership, and safety management.
They have recently viewed a [COMPANY\_1]'s job posting for a Warehouse Manager at [LOCATION] and are actively
searching for terms like ``manager,'' ``assistant,'' and ``Sales Coordination.'' Their profile highlights a background in logistics,
safety, and customer service, with recent experience at [COMPANY\_2] and [COMPANY\_3].
They are transitioning from contractor to full-time roles and have skills in lead generation, social media management, and rapid learning.

\bigskip

\textbf{Relevance Ranking:}\\
1. Warehouse Manager\\
2. Assistant Manager\\
3. Sales Consultant\\
4. Manager (general)\\
\end{tcolorbox}

\vspace{-6pt}
%{\footnotesize Emergence of a ranked sample generated by unconstrained RL. The text was produced without explicit prompting, suggesting that the resulting ranked list may serve as a potentially useful feature. Portions of the text have been redacted for privacy, indicated by black boxes. %Generation of such text without explicit prompting alludes to the fact that such a sample ranked list is potentially a useful feature. Portions of the text have been redacted for privacy, indicated by black boxes.}
\end{table}

\begin{table}[t]
\centering
\caption{Input refinement.}
\vspace{-12pt}
\caption*{\footnotesize Generated text from an input profile containing only current role and education. The model was able to infer potential skills and target roles even without explicit information. }
\label{table:refinement}
\vspace{-12pt}
\begin{tcolorbox}[myblock,width=\columnwidth,colback=gray!4,top=2pt,bottom=1pt]
\footnotesize
The member is a seasoned educator with [YEARS] years of experience, currently serving as the [ROLE] of a secondary school in  [LOCATION]. They hold a strong academic background in [SKILL] and a master’s degree in [STUDY]. Their skills include leadership, public speaking, and proficiency in Microsoft Office tools. They are likely targeting roles in education, particularly in management or curriculum development, given their experience and academic focus.
\end{tcolorbox}

\vspace{-6pt}
%{\footnotesize Generated text from an input profile that contained only current role and education. Model was able to infer likely skills and likely roles they may target despite lack of explicit information. Portions of the text have been redacted for privacy, indicated by black boxes.}
\end{table}

\begin{table}[t]
\centering
\caption{Qualitative evaluation.}
\label{table:qualitative_eval}
\vspace{-8pt}
\scalebox{0.88}{%
\begin{tabular}{lccc}
\toprule
\textbf{Model} & \textbf{Acc.} & \textbf{Cov.} & \textbf{Conc.} \\
\midrule
Pointwise String-based Model & 89.5\% & 80.7\% & 94\% \\
Pointwise Logprob-based Model & 84.7\% & 80.1\%  & 75\% \\
Listwise Model & 89\% & 78.7\%  & 96\%\\
\bottomrule
\end{tabular}
}
\vspace{-6pt}
\end{table}

\section{Qualitative Evaluation}
As depicted in Figure~\ref{fig:reward}, all three reward approaches significantly improve validation reward during the course of training. However, qualitative evaluation sheds light on differences in policies learned, as illustrated by the example in Table~\ref{table:reward_box_example}. Since listwise ranking is an easier task than pointwise action prediction, summaries from latter are predictably shorter and less granular. Moreover, listwise reward does not require the format penalty introduced in Section~\ref{subsubsect:format_penalty}, as ranking tends to be less sensitive to noise than pointwise labels. In our experiments, logprob-based rewards were particularly susceptible to reward hacking. Rewarding label probability appears to induce a `selection effect' in the output summaries, causing the model to focus largely on user activities while incorporating only a small amount of content from other sources. This is intuitive as the model prioritizes decisive features that increase prediction confidence and disregards subtler information that has only a marginal influence on log probabilities.
\subsection{Emergent Behavior}
\label{sec:emergent_behavior}

\subsubsection{Feature Discovery}
% As illustrated in Table~\ref{table:redacted_example}, the model sometimes produces ranked job samples in its generated text even without explicit prompting. Notably, this behavior emerges during reinforcement learning training when no format-based penalty is applied, indicating that the model has learned to treat ranked lists as informative structure rather than noise. This emergent capability suggests new research directions, such as explicitly eliciting ranked job preferences from users. In our work, however, we deliberately constrain the model with a format-based reward that discourages the inclusion of ranked lists in generated outputs. This design choice reflects the practical difficulty of reliably verifying such rankings for every member at scale. 
As illustrated in Table~\ref{table:redacted_example}, we observe an emergent behavior in which the actor generates an explicit ranked list of potential job matches, despite receiving no explicit instruction or prompt template requiring a ranked format. This behavior arises primarily during RL training when format-based penalties are removed. From an optimization perspective, this suggests that the model has learned to treat ranked lists as informative structure rather than noise. This suggests new research directions, such as explicitly eliciting ranked job preferences from users. In our work, however, we deliberately constrain the model with a format-based reward that discourages the inclusion of ranked lists in generated outputs. This design choice reflects the practical difficulty of reliably verifying such rankings for every member at scale. 

\subsubsection{Profile Refinement}
% We also find, surprisingly, that when user input is very short, the model refines and expands the profile to make it more useful for downstream models. For example, in Table~\ref{table:refinement}, the user had only provided their education and current title, but the model was able to infer their industry, seniority, and broad roles for which they would be suitable.
A key challenge in recommender systems is user data sparsity, particularly for cold-start or low-activity members. We find that the model exhibits a form of \textit{latent profile imputation}: when provided with sparse user inputs (e.g., only current role and education), the model leverages its pre-trained priors to infer missing contextual attributes.
For example, in Table~\ref{table:refinement}, given minimal inputs, the model infers plausible industry context, seniority signals, and candidate target roles. This implicit ``densification'' of the user representation helps bridge the gap between sparse observed data and the richer semantic features required by downstream ranking models.

\subsection{LLM as a Judge}
We adopt an LLM-as-a-judge framework to systematically evaluate the generated text along three complementary axes. For each axis, the judge yields a scalar score accompanied by a concise rationale. The evaluation dimensions are defined as follows:
\begin{compact_enum}
\item \textbf{Accuracy (Acc.)} measures factuality and groundedness. The judge checks if every atomic claim in the synopsis is entailed by one or more input sources. 
% The accuracy score is computed as the fraction of claims classified as entailed by the judge.
\item \textbf{Coverage (Cov.)} measures cross-source representativeness. This measures what percentage of input sources are reflected in information contained within the produced text. 
% The judge determines which of these units are reflected in the synopsis and returns a coverage ratio.
\item \textbf{Conciseness (Conc.)} measures brevity and non-redundancy. The judge identifies redundant or restative spans (sentences or clauses that add no materially new information) and computes a redundancy-aware conciseness score. %We track if a given line adds no new information and is simply a restatement of any of the previous lines. We also check if a line overlaps heavily with earlier lines but introduces only a small nuance or framing.
\end{compact_enum}

We use GPT-o1 to judge the summaries, requesting both a score and reasoning for the parameters described above. A small subsample is first manually annotated and used to refine the evaluation prompt. Upon running the judge, both the reasoning and the scores generated by the model are verified by human annotators.

%We use GPT-o1 to judge the summary and ask for the score as well as the reasoning for the aforementioned parameters. We first manually annotate a small subsample which is then used to iterate on the evaluation prompt. Upon running the judge,  reasoning and score are also verified by human annotators to get the final values.

Table~\ref{table:qualitative_eval} demonstrates that both the pointwise string-based and listwise models achieve high accuracy and coverage with minimal repetition. While the pointwise model slightly leads in accuracy, the listwise model excels in conciseness. 
The reward hacking observed for logprob-based model is reflected in the LLM evaluations, where it exhibits lower accuracy and reduced conciseness.

%point-wise string reward and listwise rewards both offer high accuracy, coverage and low repetition rates though the former performs better. While accuracy is largely the same for both of them, harder task results in better coverage. The reward hacking seen for logprob rewards is reflected in LLM judge evaluations as well where they suffer in accuracy and conciseness.

\subsection{User Survey}
To corroborate our offline evaluation with human judgment, we conducted a user study with 20 volunteer domain experts from LinkedIn, spanning Engineering, Product, Data Science, and Management functions. Participants rated their individual summary on a scale of 1 to 5 on two dimensions: how well it captured their job‑seeking characteristics and its factual precision. On average, the generated summaries received a score of 4.0 for content quality and 4.05 for factual accuracy. Details in Appendix section \ref{appendix:user_feedback}.

\section {Quantitative Evaluation}
\subsection{Offline Evaluation}
% The produced text was evaluated using LinkedIn's pure-text LLM-based engagement prediction model. All manually selected and hand-crafted text features were replaced with summaries from the best-performing models obtained from both pointwise and listwise rewards. We also compare it against using full text and incorporating embedding into the classification layer. As shown in Table~\ref{table:quantitative_auc_1}, adding a summary of the pointwise string-based model leads to a substantial improvement in the AUC for the prediction of pClick. We further observe that the summary can be truncated depending on online latency requirements while still providing significant improvement over the baseline. The listwise reward summary also yields a notable improvement, though it is less pronounced than the pointwise variant. We hypothesize that, since the engagement model is trained in a pointwise manner, summaries trained with pointwise reward are more informative. 

We evaluate the quality of the generated summaries using LinkedIn’s production pure-text LLM engagement predictor. Specifically, we replace the standard hand-crafted textual features with summaries produced by our best-performing policies. We compare against two baselines: (i) \textit{Full Context}, which concatenates all available raw textual sources (high latency), and (ii) \textit{Embedding}, which injects pre-computed user dense embedding vectors into the predictor’s classification layer.
As shown in Table~\ref{table:quantitative_auc_1}, the \textit{pointwise string-based} summary achieves the largest improvement in pClick AUC. The performance gain is robust to truncation: even under aggressive length constraints designed to satisfy online latency budgets, the summary consistently outperforms the baseline. While the listwise summary also yields meaningful improvements, it underperforms the pointwise variant. We hypothesize this is due to \textit{objective congruence}: the downstream engagement predictor is optimized with a pointwise binary cross-entropy objective, which is more aligned with summaries trained using pointwise reward signals.

We also evaluated the generated summaries in the retrieval layer as an additional semantic signal. This integration yields a {3\%  lift in Recall@10} without measurable degradation in relevance, suggesting that the summaries capture latent user intents that are not well represented by sparse keyword-based retrieval features.
We further apply the learned representations to improve the query rewriting module for Job Alerts. By conditioning the query generator on our user summaries, we observe consistent CTR gains in log-replay evaluation (Table~\ref{table:quantitative_auc_2}). These gains persist across both warm-start users (with explicit preferences) and cold-start users (without explicit preferences), indicating that the model can infer actionable preferences even under limited observed evidence.

% This summary feature was also used as a feature in the retrieval layer, where it led to a\textbf{ 3\% \ improvement in recall@10} without a negative impact on relevance metrics. We further evaluated the generated member text to rewrite queries in the internal job alert model. The member representation generated is used as a feature to improve derived queries for users who wish to receive job alerts. As shown in Table~\ref{table:quantitative_auc_2}, we observe estimated CTR gains under log-replay evaluation for both members with explicit preferences and members without explicit preferences. 

\subsection{Online Evaluation}
This feature has been deployed and was extensively evaluated in the ranking stage of Job Search via A/B tests. We saw significant improvements in all key engagement metrics compared to the ranker with handcrafted features as detailed in Table ~\ref{table:Online_results}. The testing was conducted online with production traffic 
and lasted two weeks. All metrics presented are statistically significant ($p\_{value}$ < 0.05).

\begin{table}[t]
\centering
\caption{Validation ROC-AUC and relative improvement.}
\label{table:quantitative_auc_1}
\vspace{-8pt}
\scalebox{0.88}{%
%\resizebox{\columnwidth}{!}{%
\begin{tabular}{lcc}
\toprule
\textbf{Model} & \textbf{Val ROC-AUC} & \textbf{Improvement} \\
\midrule
Baseline & 0.743 & -- \\
Naive prompt-tuned Summary & 0.743 & +0.0\% \\
Full Context text (750 tokens) & 0.747 & +0.54\% \\
Embedding (concat to linear layer) & 0.745 & +0.26\% \\
% SFT Summarizer & 0.746 & +0.4\% \\
Listwise Reward & 0.749 & +0.83\% \\
Pointwise Reward & 0.769 & +3.50\% \\
Pointwise Reward (first two lines) & 0.747 & +0.54\% \\
Pointwise Reward (first three lines) & 0.759 & +2.15\% \\
\bottomrule
\end{tabular}%
\vspace{-6pt}
}
\end{table}

\begin{table}[t]
\centering
\caption{Query rewriting, CTR evaluation.}
\label{table:quantitative_auc_2}
\vspace{-8pt}
\scalebox{0.88}{%
\begin{tabular}{lc}
\toprule
\textbf{Model} & \textbf{CTR Improvement} \\
\midrule
With Preferences  & +1.3\% \\
Without Preferences  & +1.1\% \\
\bottomrule
\end{tabular}
}
\vspace{-6pt}
\end{table}

\begin{table}[t]
\centering
\caption{Jobs Ranking Online A/B Test}
\label{table:Online_results}
\vspace{-8pt}
\scalebox{0.88}{%
\begin{tabular}{lc}
\toprule
\textbf{Metric} & \textbf{Relative Change} \\
\midrule
CTR  & +1.48\% \\
Job Applications  & +1.2\% \\
Session Success Rate & +0.8\%
\\
Dismiss to Apply Ratio & -1.8\% \\
Apply to Viewport Ratio & +1.02\% \\
\bottomrule
\end{tabular}
}
\vspace{-6pt}
\end{table}

\section {Deployment}
This work is deployed as a scheduled daily offline flow. Since it is computationally prohibitive to re-infer all the 1.4 billion plus members on the platform, we intelligently sample member IDs whose representations are likely to have changed since their last inference. Specifically, we include members who have updated their professional profiles, as well as those with any new job applications. In addition, we incorporate members whose `seeker score' — an aggregate indicator of job-search intent and engagement—exhibits significant day-over-day change. This signal allows us to capture both members entering an active job-search phase and those whose engagement with the search platform has meaningfully declined. The inference runs on vLLM~\cite{kwon2023efficient}, sharded over 16 H100 GPUs, allowing us to infer 25 million members in a 12-hour window on a daily basis. The produced text is subsequently persisted in a key-value store for use by downstream consumers.

\section {Conclusion}

We propose an RL-based framework that condenses heterogeneous user data into a short, interpretable textual representation that can be directly consumed by LLMs without additional training or finetuning. We study multiple reward formulations and demonstrate that engagement logs provide sufficient supervision for RL optimization. Both pointwise and listwise reward modeling produce effective representations, although logprob-based rewards are more prone to reward hacking. Representation is controlled via explicit length and format penalties. Experiments across multiple LinkedIn products demonstrate consistent gains, and the framework generalizes broadly to other domains requiring multi-source feature fusion for personalized retrieval and ranking.

% We present a RL-based framework to condense multiple sources of user data into a salient, short text representation that is interpretable and can be readily used with LLMs without requiring explicit training or finetuning. We also compare different reward formulations and demonstrate that engagement logs can serve as a sufficient signal for RL. We find that both pointwise and listwise reward modeling can be used to generate such representation, although logprob-based rewards are more susceptible to reward hacking. We control representation length by applying penalties on both length and format. We demonstrate the effectiveness of our text feature across multiple products. While our focus is on LinkedIn's job seeker products, the core framework is applicable to any industry looking to integrate features from multiple sources and leverage LLMs for personalized ranking or retrieval tasks. 

% Future work includes improving quality and representativeness of the produced feature with multiple fine-grained rewards and downstream objectives. 

% As future work, we plan to further refine the reward to improve representation quality. One such avenue is to ensure that the produced representation captures the majority of qualifications from the input sources. We also plan to augment pre-trained LLMs with downstream models to obtain representations that are both general enough and yet directly optimized for downstream tasks.

\balance

\appendix

\bibliographystyle{ACM-Reference-Format}
\bibliography{LinkedInBib}

@inproceedings{liu2025scalable,
  title={A Scalable and Efficient Signal Integration System for Job Matching},
  author={Liu, Ping and Arora, Rajat and Shi, Xiao and Le, Benjamin Hoan and Shen, Qianqi and Shen, Jianqiang and Jiang, Chengming and Zhiltsov, Nikita and Bannur, Priya and Zhu, Yidan and others},
  booktitle={Proceedings of the 31st ACM SIGKDD Conference on Knowledge Discovery and Data Mining V. 2},
  pages={4659--4669},
  year={2025}
}

@inproceedings{ren2024representation,
  title={Representation learning with large language models for recommendation},
  author={Ren, Xubin and Wei, Wei and Xia, Lianghao and Su, Lixin and Cheng, Suqi and Wang, Junfeng and Yin, Dawei and Huang, Chao},
  booktitle={Proceedings of the ACM web conference 2024},
  pages={3464--3475},
  year={2024}
}

@inproceedings{10.1145/3711896.3737029,
author = {He, Yingzhi and Liu, Xiaohao and Zhang, An and Ma, Yunshan and Chua, Tat-Seng},
title = {LLM2Rec: Large Language Models Are Powerful Embedding Models for Sequential Recommendation},
year = {2025},
isbn = {9798400714542},
publisher = {Association for Computing Machinery},
address = {New York, NY, USA},
url = {https://doi.org/10.1145/3711896.3737029},
doi = {10.1145/3711896.3737029},
booktitle = {Proceedings of the 31st ACM SIGKDD Conference on Knowledge Discovery and Data Mining V.2},
pages = {896–907},
numpages = {12},
location = {Toronto ON, Canada},
series = {KDD '25}
}

@article{luo2025recranker,
  title={Recranker: Instruction tuning large language model as ranker for top-k recommendation},
  author={Luo, Sichun and He, Bowei and Zhao, Haohan and Shao, Wei and Qi, Yanlin and Huang, Yinya and Zhou, Aojun and Yao, Yuxuan and Li, Zongpeng and Xiao, Yuanzhang and others},
  journal={ACM Transactions on Information Systems},
  volume={43},
  number={5},
  pages={1--31},
  year={2025},
  publisher={ACM New York, NY}
}

@article{tang2024self,
  title={Self-retrieval: End-to-end information retrieval with one large language model},
  author={Tang, Qiaoyu and Chen, Jiawei and Li, Zhuoqun and Yu, Bowen and Lu, Yaojie and Yu, Haiyang and Lin, Hongyu and Huang, Fei and He, Ben and Han, Xianpei and others},
  journal={Advances in Neural Information Processing Systems},
  volume={37},
  pages={63510--63533},
  year={2024}
}

@inproceedings{10.1145/3690624.3709413,
author = {Shah, Jaidev and Barjasteh, Iman and Barapatre, Amey and Forsati, Rana and Luo, Gang and Wu, Fan and Fang, Yuan and Deng, Xue and Shepard, Blake and Shah, Ronak and Yang, Linjun and Li, Hongzhi},
title = {Towards Web-scale Recommendations with LLMs: From Quality-aware Ranking to Candidate Generation},
year = {2025},
isbn = {9798400712456},
publisher = {Association for Computing Machinery},
address = {New York, NY, USA},
url = {https://doi.org/10.1145/3690624.3709413},
doi = {10.1145/3690624.3709413},
booktitle = {Proceedings of the 31st ACM SIGKDD Conference on Knowledge Discovery and Data Mining V.1},
pages = {2514–2524},
numpages = {11},
location = {Toronto ON, Canada},
series = {KDD '25}
}

@misc{zhuang2025rankr1enhancingreasoningllmbased,
      title={Rank-R1: Enhancing Reasoning in LLM-based Document Rerankers via Reinforcement Learning}, 
      author={Shengyao Zhuang and Xueguang Ma and Bevan Koopman and Jimmy Lin and Guido Zuccon},
      year={2025},
      eprint={2503.06034},
      archivePrefix={arXiv},
      primaryClass={cs.IR},
      url={https://arxiv.org/abs/2503.06034}, 
}

@article{lin2025rec,
  title={{Rec-r1: Bridging generative large language models and user-centric recommendation systems via reinforcement learning}},
  author={Lin, Jiacheng and Wang, Tian and Qian, Kun},
  journal={arXiv preprint arXiv:2503.24289},
  year={2025}
}

@article{guo2025deepseek,
  title={Deepseek-r1: Incentivizing reasoning capability in llms via reinforcement learning},
  author={Guo, Daya and Yang, Dejian and Zhang, Haowei and Song, Junxiao and Zhang, Ruoyu and Xu, Runxin and Zhu, Qihao and Ma, Shirong and Wang, Peiyi and Bi, Xiao and others},
  journal={arXiv preprint arXiv:2501.12948},
  year={2025}
}

@inproceedings{singh2024better,
  title={Better generalization with semantic ids: A case study in ranking for recommendations},
  author={Singh, Anima and Vu, Trung and Mehta, Nikhil and Keshavan, Raghunandan and Sathiamoorthy, Maheswaran and Zheng, Yilin and Hong, Lichan and Heldt, Lukasz and Wei, Li and Tandon, Devansh and others},
  booktitle={Proceedings of the 18th ACM Conference on Recommender Systems},
  pages={1039--1044},
  year={2024}
}

@inproceedings{wang2024learnable,
  title={Learnable item tokenization for generative recommendation},
  author={Wang, Wenjie and Bao, Honghui and Lin, Xinyu and Zhang, Jizhi and Li, Yongqi and Feng, Fuli and Ng, See-Kiong and Chua, Tat-Seng},
  booktitle={Proceedings of the 33rd ACM International Conference on Information and Knowledge Management},
  pages={2400--2409},
  year={2024}
}

@article{fang2025hid,
  title={HiD-VAE: Interpretable Generative Recommendation via Hierarchical and Disentangled Semantic IDs},
  author={Fang, Dengzhao and Gao, Jingtong and Zhu, Chengcheng and Li, Yu and Zhao, Xiangyu and Chang, Yi},
  journal={arXiv preprint arXiv:2508.04618},
  year={2025}
}

@inproceedings{li2024calrec,
  title={Calrec: Contrastive alignment of generative llms for sequential recommendation},
  author={Li, Yaoyiran and Zhai, Xiang and Alzantot, Moustafa and Yu, Keyi and Vuli{\'c}, Ivan and Korhonen, Anna and Hammad, Mohamed},
  booktitle={Proceedings of the 18th ACM Conference on Recommender Systems},
  pages={422--432},
  year={2024}
}

@inproceedings{zhang2024embsum,
  title={Embsum: Leveraging the summarization capabilities of large language models for content-based recommendations},
  author={Zhang, Chiyu and Sun, Yifei and Wu, Minghao and Chen, Jun and Lei, Jie and Abdul-Mageed, Muhammad and Jin, Rong and Liu, Angli and Zhu, Ji and Park, Sem and others},
  booktitle={Proceedings of the 18th ACM Conference on Recommender Systems},
  pages={1010--1015},
  year={2024}
}

@article{zhao2023embedding,
  title={Embedding in recommender systems: A survey},
  author={Zhao, Xiangyu and Wang, Maolin and Zhao, Xinjian and Li, Jiansheng and Zhou, Shucheng and Yin, Dawei and Li, Qing and Tang, Jiliang and Guo, Ruocheng},
  journal={arXiv preprint arXiv:2310.18608},
  year={2023}
}

@inproceedings{qiu2021u,
  title={U-BERT: Pre-training user representations for improved recommendation},
  author={Qiu, Zhaopeng and Wu, Xian and Gao, Jingyue and Fan, Wei},
  booktitle={Proceedings of the AAAI Conference on Artificial Intelligence},
  volume={35},
  number={5},
  pages={4320--4327},
  year={2021}
}

@inproceedings{wang2022learning,
  title={Learning supplementary NLP features for CTR prediction in sponsored search},
  author={Wang, Dong and Yan, Shaoguang and Xia, Yunqing and Salamatian, Kav{\'e} and Deng, Weiwei and Zhang, Qi},
  booktitle={Proceedings of the 28th ACM SIGKDD Conference on Knowledge Discovery and Data Mining},
  pages={4010--4020},
  year={2022}
}

@inproceedings{zhang2023twhin,
  title={Twhin-bert: A socially-enriched pre-trained language model for multilingual tweet representations at twitter},
  author={Zhang, Xinyang and Malkov, Yury and Florez, Omar and Park, Serim and McWilliams, Brian and Han, Jiawei and El-Kishky, Ahmed},
  booktitle={Proceedings of the 29th ACM SIGKDD conference on knowledge discovery and data mining},
  pages={5597--5607},
  year={2023}
}

@article{williams1992simple,
  title={{Simple statistical gradient-following algorithms for connectionist reinforcement learning}},
  author={Williams, Ronald J},
  journal={Machine learning},
  volume={8},
  number={3},
  pages={229--256},
  year={1992},
  publisher={Springer}
}

@article{sutton1999policy,
  title={{Policy gradient methods for reinforcement learning with function approximation}},
  author={Sutton, Richard S and McAllester, David and Singh, Satinder and Mansour, Yishay},
  journal={Advances in neural information processing systems},
  volume={12},
  year={1999}
}

@inproceedings{schulman2015trust,
  title={{Trust region policy optimization}},
  author={Schulman, John and Levine, Sergey and Abbeel, Pieter and Jordan, Michael and Moritz, Philipp},
  booktitle={International conference on machine learning},
  pages={1889--1897},
  year={2015},
  organization={PMLR}
}

@article{schulman2017proximal,
  title={{Proximal policy optimization algorithms}},
  author={Schulman, John and Wolski, Filip and Dhariwal, Prafulla and Radford, Alec and Klimov, Oleg},
  journal={arXiv preprint arXiv:1707.06347},
  year={2017}
}

@article{shao2024deepseekmath,
  title={{Deepseekmath: Pushing the limits of mathematical reasoning in open language models}},
  author={Shao, Zhihong and Wang, Peiyi and Zhu, Qihao and Xu, Runxin and Song, Junxiao and Bi, Xiao and Zhang, Haowei and Zhang, Mingchuan and Li, YK and Wu, Yang and others},
  journal={arXiv preprint arXiv:2402.03300},
  year={2024}
}

@article{yu2025dapo,
  title={{Dapo: An open-source llm reinforcement learning system at scale}},
  author={Yu, Qiying and Zhang, Zheng and Zhu, Ruofei and Yuan, Yufeng and Zuo, Xiaochen and Yue, Yu and Dai, Weinan and Fan, Tiantian and Liu, Gaohong and Liu, Lingjun and others},
  journal={arXiv preprint arXiv:2503.14476},
  year={2025}
}

@article{liu2025understanding,
  title={{Understanding r1-zero-like training: A critical perspective}},
  author={Liu, Zichen and Chen, Changyu and Li, Wenjun and Qi, Penghui and Pang, Tianyu and Du, Chao and Lee, Wee Sun and Lin, Min},
  journal={arXiv preprint arXiv:2503.20783},
  year={2025}
}

@inproceedings{wang2024reinforcement,
  title={Reinforcement learning-based recommender systems with large language models for state reward and action modeling},
  author={Wang, Jie and Karatzoglou, Alexandros and Arapakis, Ioannis and Jose, Joemon M},
  booktitle={Proceedings of the 47th International ACM SIGIR conference on research and development in information retrieval},
  pages={375--385},
  year={2024}
}

@article{mao2025reinforced,
  title={Reinforced prompt personalization for recommendation with large language models},
  author={Mao, Wenyu and Wu, Jiancan and Chen, Weijian and Gao, Chongming and Wang, Xiang and He, Xiangnan},
  journal={ACM Transactions on Information Systems},
  volume={43},
  number={3},
  pages={1--27},
  year={2025},
  publisher={ACM New York, NY}
}

@inproceedings{ji2024genrec,
  title={Genrec: Large language model for generative recommendation},
  author={Ji, Jianchao and Li, Zelong and Xu, Shuyuan and Hua, Wenyue and Ge, Yingqiang and Tan, Juntao and Zhang, Yongfeng},
  booktitle={European Conference on Information Retrieval},
  pages={494--502},
  year={2024},
  organization={Springer}
}

@inproceedings{kwon2023efficient,
  title={{Efficient memory management for large language model serving with pagedattention}},
  author={Kwon, Woosuk and Li, Zhuohan and Zhuang, Siyuan and Sheng, Ying and Zheng, Lianmin and Yu, Cody Hao and Gonzalez, Joseph and Zhang, Hao and Stoica, Ion},
  booktitle={Proceedings of the 29th symposium on operating systems principles},
  pages={611--626},
  year={2023}
}

@article{shani2005mdp,
  title={An MDP-based recommender system},
  author={Shani, Guy and Heckerman, David and Brafman, Ronen I},
  journal={Journal of machine Learning research},
  volume={6},
  number={Sep},
  pages={1265--1295},
  year={2005}
}

@inproceedings{moling2012optimal,
  title={Optimal radio channel recommendations with explicit and implicit feedback},
  author={Moling, Omar and Baltrunas, Linas and Ricci, Francesco},
  booktitle={Proceedings of the sixth ACM conference on Recommender systems},
  pages={75--82},
  year={2012}
}

@article{afsar2022reinforcement,
  title={Reinforcement learning based recommender systems: A survey},
  author={Afsar, M Mehdi and Crump, Trafford and Far, Behrouz},
  journal={ACM Computing Surveys},
  volume={55},
  number={7},
  pages={1--38},
  year={2022},
  publisher={ACM New York, NY}
}

@inproceedings{taghipour2007usage,
  title={Usage-based web recommendations: a reinforcement learning approach},
  author={Taghipour, Nima and Kardan, Ahmad and Ghidary, Saeed Shiry},
  booktitle={Proceedings of the 2007 ACM conference on Recommender systems},
  pages={113--120},
  year={2007}
}

@inproceedings{wang2025policy,
  title={Policy-Guided Causal State Representation for Offline Reinforcement Learning Recommendation},
  author={Wang, Siyu and Chen, Xiaocong and Yao, Lina},
  booktitle={Proceedings of the ACM on Web Conference 2025},
  pages={402--412},
  year={2025}
}

@article{lee2025gemini,
  title={Gemini embedding: Generalizable embeddings from gemini},
  author={Lee, Jinhyuk and Chen, Feiyang and Dua, Sahil and Cer, Daniel and Shanbhogue, Madhuri and Naim, Iftekhar and {\'A}brego, Gustavo Hern{\'a}ndez and Li, Zhe and Chen, Kaifeng and Vera, Henrique Schechter and others},
  journal={arXiv preprint arXiv:2503.07891},
  year={2025}
}

@inproceedings{li2023text,
  title={Text is all you need: Learning language representations for sequential recommendation},
  author={Li, Jiacheng and Wang, Ming and Li, Jin and Fu, Jinmiao and Shen, Xin and Shang, Jingbo and McAuley, Julian},
  booktitle={Proceedings of the 29th ACM SIGKDD Conference on Knowledge Discovery and Data Mining},
  pages={1258--1267},
  year={2023}
}

@article{penaloza2024tears,
  title={TEARS: Textual Representations for Scrutable Recommendations},
  author={Penaloza, Emiliano and Gouvert, Olivier and Wu, Haolun and Charlin, Laurent},
  journal={arXiv preprint arXiv:2410.19302},
  year={2024}
}

@article{gao2024end,
  title={End-to-end Training for Recommendation with Language-based User Profiles},
  author={Gao, Zhaolin and Zhou, Joyce and Dai, Yijia and Joachims, Thorsten},
  journal={arXiv preprint arXiv:2410.18870},
  year={2024}
}

@inproceedings{mysore2023editable,
  title={Editable user profiles for controllable text recommendations},
  author={Mysore, Sheshera and Jasim, Mahmood and McCallum, Andrew and Zamani, Hamed},
  booktitle={Proceedings of the 46th International ACM SIGIR Conference on Research and Development in Information Retrieval},
  pages={993--1003},
  year={2023}
}

@inproceedings{ju2025learning,
  title={Learning Universal User Representations Leveraging Cross-domain User Intent at Snapchat},
  author={Ju, Clark Mingxuan and Neves, Leonardo and Kumar, Bhuvesh and Collins, Liam and Zhao, Tong and Qiu, Yuwei and Dou, Qing and Zhou, Yang and Nizam, Sohail and Ozturk, Rengim Aykan and others},
  booktitle={Proceedings of the 48th International ACM SIGIR Conference on Research and Development in Information Retrieval},
  pages={4345--4349},
  year={2025}
}

@inproceedings{geng2022recommendation,
  title={Recommendation as language processing (rlp): A unified pretrain, personalized prompt \& predict paradigm (p5)},
  author={Geng, Shijie and Liu, Shuchang and Fu, Zuohui and Ge, Yingqiang and Zhang, Yongfeng},
  booktitle={Proceedings of the 16th ACM conference on recommender systems},
  pages={299--315},
  year={2022}
}

@inproceedings{zhang2024recgpt,
  title={RecGPT},
  author={Zhang, Yabin and Yu, Wenhui and Zhang, Erhan and Chen, Xu and Hu, Lantao and Jiang, Peng and Gai, Kun},
  booktitle={International Conference on Database Systems for Advanced Applications},
  pages={233--250},
  year={2024},
  organization={Springer}
}

@inproceedings{sheng2025hybridflow,
  title={{Hybridflow: A flexible and efficient rlhf framework}},
  author={Sheng, Guangming and Zhang, Chi and Ye, Zilingfeng and Wu, Xibin and Zhang, Wang and Zhang, Ru and Peng, Yanghua and Lin, Haibin and Wu, Chuan},
  booktitle={Proceedings of the Twentieth European Conference on Computer Systems},
  pages={1279--1297},
  year={2025}
}

@article{kaelbling1996reinforcement,
  title={{Reinforcement learning: A survey}},
  author={Kaelbling, Leslie Pack and Littman, Michael L and Moore, Andrew W},
  journal={Journal of artificial intelligence research},
  volume={4},
  pages={237--285},
  year={1996}
}

@book{sutton1998reinforcement,
  title={{Reinforcement learning: An introduction}},
  author={Sutton, Richard S and Barto, Andrew G and others},
  volume={1},
  number={1},
  year={1998},
  publisher={MIT press Cambridge}
}

@article{watkins1992q,
  title={Q-learning},
  author={{Watkins, Christopher JCH and Dayan, Peter}},
  journal={Machine learning},
  volume={8},
  number={3},
  pages={279--292},
  year={1992},
  publisher={Springer}
}

@article{mnih2013playing,
  title={{Playing atari with deep reinforcement learning}},
  author={Mnih, Volodymyr and Kavukcuoglu, Koray and Silver, David and Graves, Alex and Antonoglou, Ioannis and Wierstra, Daan and Riedmiller, Martin},
  journal={arXiv preprint arXiv:1312.5602},
  year={2013}
}

@article{dash1997feature,
  title={Feature selection for classification},
  author={Dash, Manoranjan and Liu, Huan},
  journal={Intelligent data analysis},
  volume={1},
  number={1-4},
  pages={131--156},
  year={1997},
  publisher={Elsevier}
}

@inproceedings{huang2020embedding,
  title={Embedding-based retrieval in facebook search},
  author={Huang, Jui-Ting and Sharma, Ashish and Sun, Shuying and Xia, Li and Zhang, David and Pronin, Philip and Padmanabhan, Janani and Ottaviano, Giuseppe and Yang, Linjun},
  booktitle={KDD'20},
  pages={2553--2561},
  year={2020}
}

@inproceedings{ziegler2019fine,
  title={Fine-tuning language models from human preferences},
  author={Ziegler, Daniel M and Stiennon, Nisan and Wu, Jeffrey and Brown, Tom B and Radford, Alec and Amodei, Dario and Christiano, Paul and Irving, Geoffrey},
  booktitle={Advances in Neural Information Processing Systems (NeurIPS)},
  volume={32},
  year={2019}
}

@inproceedings{gao2023scaling,
  title={Scaling Laws for Reward Model Overoptimization},
  author={Gao, Leo and Schulman, John and Hilton, Jacob and Durmus, Esin},
  booktitle={International Conference on Machine Learning (ICML)},
  year={2023}
}

@inproceedings{ng1999policy,
  title={Policy invariance under reward transformations: Theory and application to reward shaping},
  author={Ng, Andrew Y and Harada, Daishi and Russell, Stuart},
  booktitle={Proceedings of the 16th International Conference on Machine Learning (ICML)},
  pages={278--287},
  year={1999}
}

@inproceedings{lyu2025dv365,
  title={DV365: Extremely Long User History Modeling at Instagram},
  author={Lyu, Wenhan and Tyagi, Devashish and Yang, Yihang and Li, Ziwei and Somani, Ajay and Shanmugasundaram, Karthikeyan and Andrejevic, Nikola and Adeputra, Ferdi and Zeng, Curtis and Singh, Arun K and others},
  booktitle={Proceedings of the 31st ACM SIGKDD Conference on Knowledge Discovery and Data Mining V. 2},
  pages={4717--4727},
  year={2025}
}

@inproceedings{ye2025applying,
  title={Applying Large Language Model For Relevance Search In Tencent},
  author={Ye, Dezhi and Liu, Jie and Hu, Junwei and Fan, Jiabin and Tian, Bowen and Liang, Haijin and Ma, Jin},
  booktitle={Proceedings of the 31st ACM SIGKDD Conference on Knowledge Discovery and Data Mining V. 2},
  pages={5171--5181},
  year={2025}
}

@article{huang2025towards,
  title={Towards agentic recommender systems in the era of multimodal large language models},
  author={Huang, Chengkai and Wu, Junda and Xia, Yu and Yu, Zixu and Wang, Ruhan and Yu, Tong and Zhang, Ruiyi and Rossi, Ryan A and Kveton, Branislav and Zhou, Dongruo and others},
  journal={arXiv preprint arXiv:2503.16734},
  year={2025}
}

@inproceedings{shah2025towards,
  title={Towards Web-scale Recommendations with LLMs: From Quality-aware Ranking to Candidate Generation},
  author={Shah, Jaidev and Barjasteh, Iman and Barapatre, Amey and Forsati, Rana and Luo, Gang and Wu, Fan and Fang, Yuan and Deng, Xue and Shepard, Blake and Shah, Ronak and others},
  booktitle={Proceedings of the 31st ACM SIGKDD Conference on Knowledge Discovery and Data Mining V. 1},
  pages={2514--2524},
  year={2025}
}

@inproceedings{palumbo2025you,
  title={You Say Search, I Say Recs: A Scalable Agentic Approach to Query Understanding and Exploratory Search at Spotify},
  author={Palumbo, Enrico and Isaksson, Marcus and Tamborrino, Alexandre and Movin, Maria and Dincu, Catalin and Vardasbi, Ali and Nikeshkin, Lev and Gorobets, Oksana and Nyman, Anders and Newdick, Poppy and others},
  booktitle={Proceedings of the Nineteenth ACM Conference on Recommender Systems},
  pages={1117--1121},
  year={2025}
}

@article{wei2022chain,
  title={Chain-of-thought prompting elicits reasoning in large language models},
  author={Wei, Jason and Wang, Xuezhi and Schuurmans, Dale and Bosma, Maarten and Xia, Fei and Chi, Ed and Le, Quoc V and Zhou, Denny and others},
  journal={Advances in neural information processing systems},
  volume={35},
  pages={24824--24837},
  year={2022}
}

 % \clearpage
% \begin{acks}
% \end{acks}

%%


% \section{Reward Model Selection}
% \label{appendix:reward_model_selection}
% We adopt Qwen3-32B as the primary reward model due to its strong performance in~\cite{yang2025qwen3} and high rankings on public benchmarks. Member raw information, including profiles, resumes, activities, and job search queries, is used as input for evaluation. We assess both pointwise (binary classification) and listwise (job ranking) prediction tasks with Qwen3-32B and o1~\cite{jaech2024openai}. For pointwise prediction, Qwen3-32B achieves $0.72$ accuracy compared to $0.74$ for o1. For listwise prediction, Qwen3-32B achieves a \gls{ndcg} score of $0.7604$ while o1 achieves $0.7256$. Overall, these results support the effectiveness of Qwen3-32B.

% In addition to evaluating Qwen3-32B, this process provides several practical benefits. It enables the identification of reward score ranges and the estimation of performance ceilings during \gls{rl} training. It also guides prompt tuning prior to \gls{rl} training, thereby supporting more efficient model development.
\section*{Appendix}

\section{Ethical Consideration}
This study uses de-identified member information to train LLMs while ensuring strict adherence to ethical research practices. No personally identifiable information (PII) was collected, stored, or used in the training or evaluation processes. The methodology was designed to minimize the risks of re-identification, bias, or misuse, and follows principles of fairness, accountability, and transparency.
\section {User Feedback}
\label{appendix:user_feedback}
Participants rated each summary on a 1–5 scale along two axes: (i) how well the summary captured their key characteristics and (ii) factual accuracy. A score of 5 indicated that the summary represented them extremely well and was fully accurate, whereas a score of 1 indicated that the summary omitted major characteristics or contained substantial factual errors.

\begin{figure}[htb]
    \centering
    \includegraphics[width=0.34\textwidth]{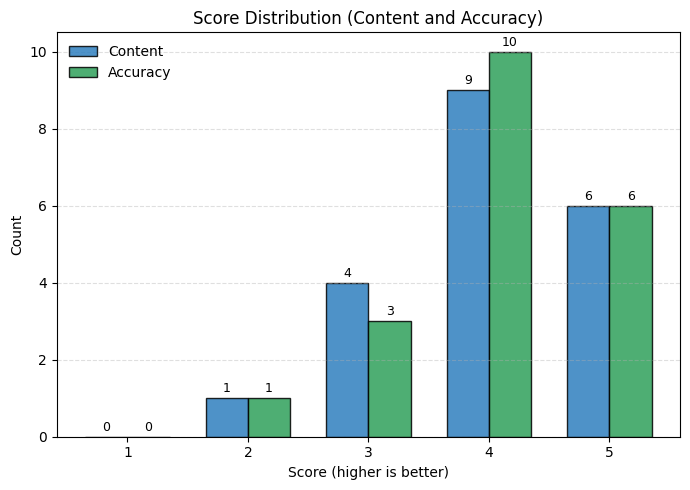}
    \caption{Distribution of human ratings for content and accuracy (n=20)}
    \label{fig:survey}
\end{figure}

Figure \ref{fig:survey} shows the distribution of scores. On average, the generated summaries received a score of 4.0 for content quality and 4.05 for accuracy.

To illustrate the range of participant feedback—including both critical and highly positive responses—we include selected excerpts below:
\begin{quote}
\textbf{(Content 4, Accuracy 4)} \\
\emph{``Overall it is an accurate summary. But the first blurb is missing that my
governance experience is in the e-commerce domain. It’s minor, but my profile
specifies governance at \textless company\textgreater{}, which may differ from
governance in other industries.''}
\end{quote}

\begin{quote}
\textbf{(Content 5, Accuracy 5)} \\
\emph{``I like how it captures my recent background while giving appropriate
weight to my recent queries as well.''}
\end{quote}

\begin{quote}
\textbf{(Content 3, Accuracy 4)} \\
\emph{``If the target job is MLE, I would prefer the summary to highlight my MLE
background (e.g., predictive modeling) and downplay less relevant aspects such as
general data analysis.''}
\end{quote}

\section{Prompt Templates}
\label{appendix:prompt_templates}
This section presents  prompt templates used by the reward model, with example synthetic inputs (see Table~\ref{table:reward_class_template} and Table~\ref{table:reward_rank_template}).

% \section{Samples from Ramped Member Text Representation Model}
% Table~\ref{table:sample_generations_combined} provides additional examples to show differences in generated member text resulting from pointwise versus listwise rewards. The source profile were anonymized by replacing entities with their surrogates.

\begin{table}[H]
\caption{Pointwise reward prompt template along with a synthetic example input for demonstration purposes.}

%\vspace{-8pt}
\centering
\small   % or \footnotesize
\begin{tcolorbox}[enhanced, width=\linewidth, colback=white]
\textbf{Pointwise Reward Prompt Template:} \\[0.3em]
\texttt{<|im\_start|>system} \\
You are an assistant tasked with predicting whether a member would apply to a specific job posting, based solely on the member's summary. \\ \\
The member summary may include high-level information such as background, interests, career goals, or inferred intent. Use this summary to evaluate the relevance and appeal of the job posting to the member. \\ \\
Interpretation Guidelines:
\\ - apply: High interest or strong alignment with the member's background and intent.
\\ - not apply: Low interest or weak alignment with the member's background and intent. \\ \\
Output Instructions:
\\ - Output EXACTLY ONE WORD: yes if the member would apply, no otherwise.
\\ - Do NOT include any explanations, formatting, or punctuation.
\\ - Your output must be strictly yes or no \\
\texttt{<|im\_end|>} \\ \\
\texttt{<|im\_start|>user} \\
Input: \\
- Member Summary: \\
\{member\_summary\} \\ \\
- Job Posting: \\
\{job\_description\} \\
\texttt{<|im\_end|>}
\bigskip\hrule\bigskip
\textbf{Synthetic Example Input:} \\[0.3em]
- Member Summary: \\
The member is seeking roles in product management and operations, with a focus on process optimization, cross-functional team coordination, and project strategy. They have a solid background in operational planning, stakeholder communication, and performance analysis, and are likely interested in positions that match their experience in management and process improvement. \\ \\ 
- Job Posting: \\
This is a Product Operations Manager position at LinkedIn.
\end{tcolorbox}
\label{table:reward_class_template}
\end{table}

\begin{table}[H]
\caption{Listwise reward prompt template along with a synthetic example input for demonstration purposes.}
\centering
\small   % or \footnotesize
\begin{tcolorbox}[enhanced, width=\linewidth, colback=white]
\textbf{Listwise Reward Prompt Template:} \\[0.3em]
\texttt{<|im\_start|>system} \\
You are an assistant tasked with ranking a list of 5 new job postings based on how much interest a member would have in each job, based on their provided member summary. \\ \\
The member summary may include high-level information such as background, interests, career goals, or inferred intent. Use this summary to rank the job postings from most to least relevant. \\ \\
Interpret the member's interest based on the following signals:
\\ - Jobs that are most relevant and strongly aligned with the member's goals should be ranked at the top.
\\ - Jobs that are moderately interesting should be ranked in the middle.
\\ - Jobs that are least relevant or poorly aligned should be ranked at the bottom. \\ \\
Output Instructions:
\\ - The final answer in the following format: [job\_index\_0, job\_index\_1, job\_index\_2, job\_index\_3, job\_index\_4]
\\ - The job indices must be integers corresponding to the input list, ordered from most to least relevant.
\\ - The list must include all job indices from the input, with no duplicates and no missing values.
\\ - Do not include any additional explanation or text outside the required format.\\
\texttt{<|im\_end|>} \\ \\
\texttt{<|im\_start|>user} \\
Input: \\
- Member Summary: \\
\{member\_summary\} \\ \\
- 5 New Job Postings: \\
\{job\_descriptions\} \\
\texttt{<|im\_end|>}
\bigskip\hrule\bigskip
\textbf{Synthetic Example Input:} \\[0.3em]
- Member Summary: \\
The member is seeking roles in product management and operations, with a focus on process optimization, cross-functional team coordination, and project strategy. They have a solid background in operational planning, stakeholder communication, and performance analysis, and are likely interested in positions that match their experience in management and process improvement. \\ \\ 
- 5 New Job Postings: \\
Job 0: This is a Product Operations Manager position at LinkedIn. \\
Job 1: This is a Graphic Designer position at Salesforce. \\
Job 2: This is a Senior Product Strategy Manager position at Meta. \\
Job 3: This is a Front-End Software Engineer position at Google. \\
Job 4: This is an Operations Excellence Lead position at Amazon.
\end{tcolorbox}
\label{table:reward_rank_template}
\end{table}

\end{document}